# Relationship between work-family balance, employee well-being, and job performance

Spanish title: Relación entre conciliación trabajo-familia, bienestar del empleado y desempeño laboral

*José Aurelio Medina-Garrido, José María Biedma-Ferrer and Antonio Rafael Ramos-Rodríguez*
INDESS. Universidad de Cádiz, Spain

## Abstract

**Purpose**

To assess the impact of the existence of and access to different work-family policies on employee well-being and job performance.

**Design/methodology/approach**

Hypothesis testing was performed using a structural equation model based on a PLS-SEM approach applied to a sample of 1,511 employees of the Spanish banking sector.

**Findings**

The results obtained demonstrate that the existence and true access to different types of work-family policies such as flexible working hours (flexi-time), long leaves, and flexible work location (flexi-place) are not directly related to job performance, but indirectly so, when mediated by the well-being of employees generated by work-family policies. In a similar vein, true access to employee and family support services also has an indirect positive impact on job performance mediated by the well-being produced. In contrast, the mere existence of employee and family support services does not have any direct or indirect effect on job performance.

**Originality/value**

This study makes a theoretical and empirical contribution to better understand the impact that of the existence of and access to work-family policies on job performance mediated by employee well-being. In this sense, we posited and tested an unpublished theoretical model where the concept of employee well-being gains special relevance at academic and organizational level due to its implications for human resource management.

**Keywords**: Work-family balance, job performance, employee well-being, human resource management.

**(Abstract in Spanish)**

**Resumen**
**Propósito**
Este trabajo analiza los efectos de la existencia y accesibilidad de diferentes tipos de políticas trabajo-familia (WFP) sobre el bienestar y el desempeño laboral.

**Diseño/Metodología/Enfoque**
Para contrastar las hipótesis propuestas se aplicó un modelo de ecuaciones estructurales, utilizando el enfoque PLS-SEM, a una muestra de 1.511 trabajadores del sector financiero español.

**Resultados**
Los resultados del análisis muestran que la existencia y accesibilidad de las WFP relativas a flexibilidad temporal, permisos largos y el lugar de trabajo no producen directamente mejoras en el desempeño, pero sí indirectamente a través del bienestar laboral que dichas políticas generan. Del mismo modo, la accesibilidad de las WFP relativas a servicios de apoyo al empleado y a su familia tiene también un efecto positivo indirecto sobre el desempeño, mediado por el bienestar generado. Sin embargo, la mera existencia de servicios de apoyo no incide ni directa ni indirectamente sobre el desempeño.

**Originalidad/Valor**
Este trabajo constituye una novedosa aportación teórica y empírica sobre el impacto de la existencia y accesibilidad de las WFP en el desempeño, considerando el papel mediador del bienestar laboral en esta relación. En este sentido, se propone y comprueba empíricamente un modelo teórico inédito en la literatura, en el que el concepto de bienestar laboral cobra especial relevancia tanto a nivel académico como por sus implicaciones prácticas para la dirección.

**Palabras clave**: Conciliación trabajo-familia, rendimiento laboral, bienestar laboral, gestión de recursos humanos.

## 1. Introduction

As a consequence of the profound changes taking place in families and the labour market, balancing work and family is an increasing workforce demand (Mercure & Mircea, 2010). In the last decades, labour conditions have changed; long working hours make it difficult to meet family responsibilities, and job insecurity has increased (Rhnima et al, 2014). Similarly, changes have also occurred in families, such as the increase in dual-earner households and in the number of families with dependent family members or single parent families. In this context, when work interferes with family, work-to-family conflicts (WFC) arise (Lin, 2013; Greenhaus & Beutell, 1985).

Organizational efforts to improve employee well-being (EWB) through the development of work-family policies may help solve this problem. Work-family policies include work practices aimed at balancing work, family (López-Ibor et al, 2010), and personal demands (Felstead et al, 2002).

Evidence has been gathered by researchers of a positive relationship between work-family policies and job performance in socially supportive organizations (Biedma-Ferrer & Medina-Garrido, 2014; van Steenbergen & Ellemers, 2009; Anderson et al, 2002). In line with these findings, *the objective of this study* was to analyse the impact that different work-family policies have on job performance. As such, we designed a model for identifying the effects that different types of work-family policies have on EWB and, ultimately, on job performance. However, according to some authors, the mere existence of work-family policies is not enough for employees to attain a work-family balance (Yeandle et al, 2002; Budd & Mumford, 2005). Based on this finding, a distinction was made in our study between the existence and employees' awareness of work-family policies, and true access to these policies.

This study provides added value to the existing literature on work-family policies in various forms. First, the effects that the existence and employees' awareness of work-family policies have on job performance were separately assessed from the effects that the actual uptake of work-family policies without reprisals have on job performance (Baxter & Chesters, 2011; McDonald et al, 2005). In the existing literature, a distinction is not made between the adoption and the actual implementation of work-family policies (McDonald et al, 2005). As stated above, it is not enough that work-family policies are available, but employees must be aware of their existence and be provided true access to them (Budd y Mumford, 2005; Yeandle et al, 2002) without reprisals (Gray & Tudball, 2002; Bond, 2004). True access to work-family policies requires a supportive work-life organizational culture (Las Heras et al, 2015; Sivatte & Guadamillas, 2014). Secondly, we present an unpublished model for determining the relationship between the existence of and access to work-family policies and job performance as mediated by EWB. In third place, the probability that WFC arise and work-family policies are implemented to solve them may differ across sectors (Allen et al, 2015). It would be interesting to analyse the impact of work-family policies on job performance in the sectors such as the banking sector where employees are subject to more pressure in terms of results (Burke, 2009; Rosso, 2008), as they are more vulnerable to WFC and consequently more urgently require the implementation of work-family policies. The contribution of this study is that it examines the relationship between work-family policies and job performance in the banking sector, a scarcely studied sector (e.g., van Steenbergen & Ellemers, 2009). Finally, the decision to focus our study on the banking sector was based on the dramatic impact that the economic and financial crisis had on this sector in Spain. The crisis was accompanied by major restructuring and downsizing, offices closing (Alamá et al, 2015; Maudas, 2011), and increased working hours and pressure exerted on employees in a traditionally stressful sector (Ariza-Montes et. al, 2013). An additional effect of the financial crisis is that organizations now devote fewer resources to the implementation of work-family policies (Mihelič & & Tekavčič, 2014).

This study is structured in six sections, as follows: In Section 2, we expose the theoretical fundaments of the hypotheses formulated in this study. Section 3 describes the empirical study performed and the methodology used. Section 4 provides an analysis of the results. Section 5 comprises a discussion of results and their implications for human resource management. Finally, the conclusions, limitations of the study, and future research lines are described in Section 6.

## 2. Literature review and hypotheses

The literature provides evidence of the difficulties that employees face in balancing work and life (Watis et al, 2013; Keene & Quadagno, 2004). The occurrence of WFC may negatively affect job satisfaction, increase turnover intent, and cause more stress and even depression among other deleterious effects (Sánchez-Vidal et al, 2011; Steinmetz et al, 2008). It seems reasonable that organizations implement policies aimed at solving WFC. In this sense, studies have documented that the availability of work-family policies reduces the probability of WFC arising (Sivatte & Guadamillas, 2014).

Researchers have concluded that the lack of work-family policies may compromise EWB (Devi & Nagini, 2013; Hughes & Bozionelos, 2007). In addition, work-family policies can improve job performance (Jyothi & Jyothi, 2012; Kanwar et al, 2009), job satisfaction and organizational commitment (Jyothi & Jyothi, 2012; Poelmans & Caligiuri, 2008), where gender has a moderating influence (Cloninger et al, 2015). However, as stated above, the mere existence of work-family policies does not imply that employees widely use these policies (McDonald et al, 2005). Thus, employees must be aware of the work-family policies available in their organization (Budd & Mumford, 2005; Yeandle et al, 2002) and perceive that they can use them freely without reprisals (Gray & Tudball, 2002; Bond, 2004). Evidence has been provided that employees do not always know that they can use work-family policies (Yeandle et al, 2002), and they are seldom familiar with such policies adopted in their organization (Budd & Mumford, 2005). Furthermore, the fact that work-family policies exist in an organization does not necessarily mean that their employees perceive they can use them freely (Gray y Tudball, 2002; Bond, 2004). Employees are reluctant to use work-family policies for a number of reasons: missing out on the chance to be promoted, being regarded as poorly committed to the organization, or being worried about losing their job (Baxter & Chesters, 2011; Chinchilla et al, 2003), the latter increasing as the age of the employee decreases (Buonocore et al, 2015). Therefore, employees should perceive that their superiors are supportive when it comes to using work-family policies (Hammer et al, 2009; Sivatte & Guadamillas, 2014). Reducing the incidence of WFC involves making employees perceive they can use work-family policies freely, which requires a work-family supportive organizational culture (Las Heras et al, 2015; Sivatte & Guadamillas, 2014). According to the theory of perceived support, organizational work-family support is defined as the perception that one's employer provides helpful social support, regardless of the organizational policies available (Kinnunen et al, 2005).

The hypotheses formulated take into account the two aforementioned dimensions of work-family policies: on the one hand, the existence and employees' awareness of such policies and, on the other hand, employees' true access to work-family policies without suffering any negative consequences for their careers.

Following the recommendations by Pasamar & Valle (2011), work-family policies were classified into four groups (see Table 1): (1) Flexibility in working time (flexi-time); (2) flexibility through long paid and unpaid leaves; (3) flexibility in work location (flexi-place); and (4) employee and family support services.

Table 1. WFP resources by groups

| Groups | WFP |
|---|---|
| 1. Flexibility in working time | 1. Adapting the duration and distribution of working hours: continuous working day, breaks and working time flexibility.<br>2. Reduction in work hours to care for children and family members (part-time work).<br>3. Compressed workweek.<br>4. Taking holidays out of the regular vacation period.<br>5. Breastfeeding leaves.<br>6. Other working time flexibility arrangements. |
| 2. Long paid and unpaid leaves | 7. Leave to care for a hospitalized family member.<br>8. Leave to take a family member to a health center to receive medical assistance.<br>9. Paid leave for sickness of a family member.<br>10. Compressed breastfeeding leave (in days).<br>11. Leave for international adoption.<br>12. Leave to undergo a treatment of assisted reproduction.<br>13. Unpaid leaves (to care for children and dependent relatives).<br>14. Leave for personal reasons.<br>15. Unpaid additional holidays.<br>16. Other paid and unpaid leaves. |
| 3. Flexibility in the location of work | 17. Teleworking<br>18. Videoconferencing (except for teleworking).<br>19. Transfer to a location nearer the family home. |
| 4. Employee and family support services | 20. Workplace nurseries.<br>21. Childcare allowances.<br>22. Allowances for employees with child or elder care responsibilities.<br>23. Counselling on childcare services, schools, nursing homes for elderly and disabled people, etc.<br>24. Work and family support services for employees and their families: psychological, legal, financial support, etc.<br>25. Training in time and stress management<br>26. Counselling services on the WFP available<br>27. Other employee and family support services |

The availability of work-family policies may have positive effects on certain employee psychological aspects and attitudes. Accordingly, our hypotheses posit that the existence of and access to work-family policies (see Table 1) are related to EWB and job performance.

*2.1. Work-family policies and job performance*

The literature suggests that WFCs negatively affect job performance (Johns, 2011; Sánchez-Vidal et al, 2011; van Steenbergen & Ellemers, 2009). In a WFC, the employee perceives that a situation is either unfair or threatens his/her well-being, which causes stress and negative affectivity (Matta et al, 2014; Blanch & Aluja, 2012). These factors are predictors of a counterproductive work behaviour in the form of purposeful failure to perform job tasks effectively, sabotage, verbal and physical abuse and even theft (Penney & Spector, 2005; Spector et al, 2006). Long working hours, irrational working time and limited access to part-time work hours may be perceived by employees as unfair situations that can generate a WFC and result in poor job performance (Macinnes, 2005; Ahn, 2005; Lapierre & Allen, 2006). In contrast, evidence has been gathered by researchers of a positive relationship between work-family policies and job performance in socially supportive organizations (van Steenbergen & Ellemers, 2009; Anderson et al, 2002). When organizations provide resources that help relieve work-family strain, productivity increases (Estes et. al, 2007; Swody & Powell, 2007).

Based on the relationship between work-family policies and job performance, and taking into account both the existence of and access to different types of work-family policies (see Table 1), we postulated the following hypotheses:

> H1.1: The more strongly an employee perceives that work-family policies based on working time flexibility exist, the better his/her job performance.
>
> H1.2: The more strongly an employee perceives that work-family policies based on long paid and unpaid leaves exist, the better his/her job performance.
>
> H1.3: The more strongly an employee perceives that work-family policies based on flexibility in the work location exist, the better his/her job performance.

H1.4: The more strongly an employee perceives that work-family policies based on employee and family support services exist, the better his/her job performance.

H2.1: The more strongly an employee perceives that work-family policies based on working time flexibility are accessible, the better his/her job performance.

H2.2: The more strongly an employee perceives that work-family policies based on long paid and unpaid leaves are accessible, the better his/her job performance.

H2.3: The more strongly an employee perceives that work-family policies based on flexibility in the work location are accessible, the better his/her job performance.

H2.4: The more strongly an employee perceives that work-family policies based on employee and family support services are accessible, the better his/her job performance.

*2.2. Work-family policies and EWB*

A number of definitions of well-being have been proposed in the literature (Wright, 2014; Cropanzano & Wright, 2014; Edgar et al, 2015; van Steenbergen & Ellemers, 2009), among which EWB stands out (Zheng et al, 2015a; Ilies et al, 2007). When referring to EWB, the most common concept –the one chosen for this study– is psychological (Edgar et al, 2015) or emotion-based EWB (Wright, 2014). Bakker & Oerlemans (2011) define EWB as the prevalence of pleasant emotions (e.g. joy or happiness) over negative emotions (e.g. sadness or anger) as a result of job satisfaction.

The fact that EWB is positively or negatively affected by the experiences of employees suggests that work-family policies relate to EWB (Lucia-Casademunt et al, 2013). In this sense, work-family policies such as working time flexibility have been reported to have positive effects on EWB (Lewis, 2010). Therefore, whereas WFC may have a negative impact on EWB (Lapierre & Allen, 2006), work-family policies might be positively related to EWB (Zheng et al, 2015b; Biedma-Ferrer & Medina-Garrido, 2014).

Based on the potential relationship between work-family policies and EWB, we postulated that the existence of and access to different types of work-family policies (see Table 1) may influence EWB:

> H3.1: The more strongly an employee perceives that work-family policies based on working time flexibility exist, the better his/her EWB.
>
> H3.2: The more strongly an employee perceives that work-family policies based on long paid and unpaid leaves exist, the better his/her EWB.
>
> H3.3: The more strongly an employee perceives that work-family policies based on flexibility in the work location exist, the better his/her EWB.
>
> H3.4: The more strongly an employee perceives that work-family policies based on employee and family support services exist, the better his/her EWB.
>
> H4.1: The more strongly an employee perceives that work-family policies based on working time flexibility are accessible, the better his/her EWB.
>
> H4.2: The more strongly an employee perceives that work-family policies based on long paid and unpaid leaves are accessible, the better his/her EWB.
>
> H4.3: The more strongly an employee perceives that work-family policies based on flexibility in the work location are accessible, the better his/her EWB.
>
> H4.4: The more strongly an employee perceives that work-family policies based on employee and family support services are accessible, the better his/her EWB.

*2.3. EWB and job performance*

Evidence has been provided of a positive relationship between EWB and job performance (Zheng et al, 2015a; Lyubomirsky et al, 2005). According to the review performed by Wright (2014), emotion-based EWB seems to be positively related to job performance (Wright & Cropanzano, 2000; Wright et al, 2007). Emotion-based EWB is attained by an employee when s/he experiences psychological well-being in the form of lack of stress and emotional burnout, and positive affectivity (Wright, 2014). Positive affectivity involves an employee being enthusiastic, active, and alert, and it is a predictor of willingness to solve conflicts, optimism, creativity, and organizational commitment (Choi & Lee, 2014).

Based on the EWB that the work-family policies described in Table 1 can generate, and its influence on job performance, the following hypotheses were formulated:

> H5.1: The higher the EWB generated by work-family policies based on working time flexibility, the better the job performance.
>
> H5.2: The higher the EWB generated by work-family policies based on long paid and unpaid leaves, the better the job performance.
>
> H5.3: The higher the EWB generated by work-family policies based on flexibility in the work location, the better the job performance.
>
> H5.4: The higher the EWB generated by work-family policies based on employee and family support services, the better the job performance.

On the basis of the hypotheses proposed, Figure 1 displays a model of the potential positive relationship between the existence of and access to work-family policies and job performance either directly or mediated by EWB. This model also proposes a direct positive relationship between EWB and job performance.

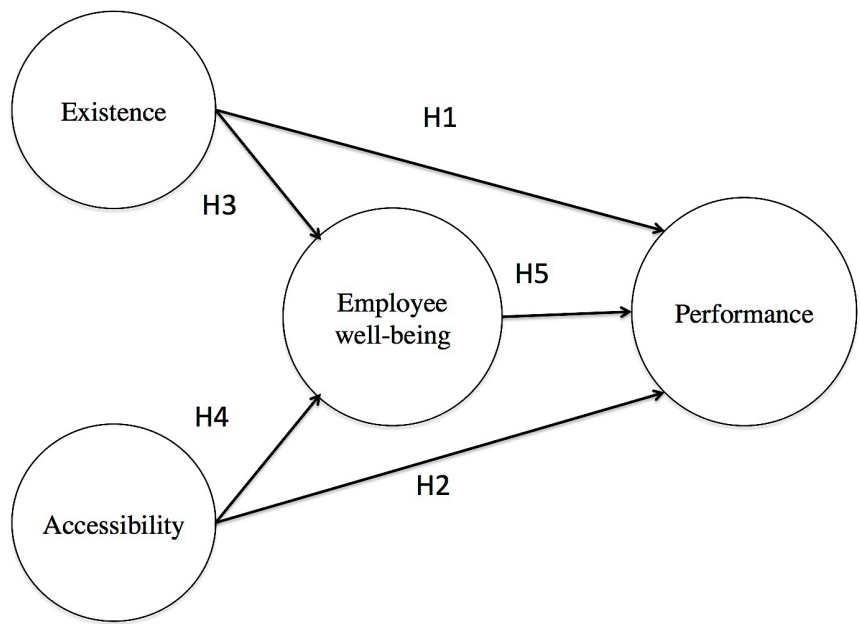

Figure 1. Theoretical model and hypotheses

As stated in our hypotheses, this model will be tested for each of the work-family policies described in Table 1.

**3. Methods**

*3.1. Sample and data collection*

Fieldwork was focused on the banking sector in Spain, a sector with a total of 192,265 employees. Data were collected using a closed-question, self-administered questionnaire. Respondents received an e-mail with a link to the web-based questionnaire. Pre-testing was conducted to improve the efficacy and design of the questionnaire, prevent ambiguity, and improve measurements. Data were collected from three organizations between July and November 2014. Organization representatives were invited to participate in the study and informed that the final aggregated results would be provided to them once they were available. A total of 1,565 questionnaires were returned by respondents, who were representative of the entire Spanish population in terms of geography. Fifty-four questionnaires were withdrawn from analysis due to incomplete or missing data. Of the resulting 1,511 effective respondents, 42.4% were female and 75.9% had child or elder care responsibilities that required work-family balance. The mean age was 43.7 years (SD 8.9), and mean seniority in the company was 18.7 years (SD 11.2).

*3.2. Measurements*

The measurement model describes the linkage between theoretical constructs and empirical data (Fornell, 1982). Two types of relationships are identified in this model: the common latent construct model (reflective indicators), i.e. when indicators represent an unobserved theoretical construct to which they are related; and the aggregated latent construct model (formative indicator), where construct modelling is based on indicators or measurements. Mackenzie et al (2005) proposed four questions for distinguishing common latent constructs from aggregated ones: (1) what is the direction of causality between the construct and its indicators? (2) Are construct indicators interchangeable in conceptual terms? (3) Are indicators correlated among themselves? And, (4) do all indicators have the same construct antecedents and consequences?

The application of these criteria to all constructs justifies the use of the reflective model, since all indicators are representations of the unobserved theoretical construct they reflect, they share a common theme, are interchangeable, and are strongly correlated.

Except for job performance, multiple indicators based on respondents' rating of a series of statements on a 5-point Likert scale were used, where 1 was "strongly disagree" and 5 was "strongly agree". Firstly, to assess the existence of work-family policies in an organization, an adapted version of the Families and Work Institute (2012a, 2012b) scale was used. The adapted scale consisted of five indicators of respondents' perceptions. Thus, respondents were asked whether work-family policies were available, whether information on work-family policies was provided to them by the organization, whether they were aware of the work-family policies, whether they knew of someone who had used them and, finally, whether they had ever used work-family policies (e.g. "I have the working time flexibility I need to meet my personal and family responsibilities".) The respective alpha coefficients for each group of work-family policies were: $\alpha_1=0.836$, $\alpha_2=0.800$, $\alpha_3=0.884$, $\alpha_4=0.924$ for flexi-time, long paid and unpaid leaves, flexi-place, and family support services, respectively.

To measure the variable *Accessibility,* a 2-item scale ($\alpha_1=0.819$, $\alpha_2=0.792$, $\alpha_3=0.851$, $\alpha_4=0.885$) based on the contributions by Anderson et al (2002) and the Families and Work Institute (2012a, 2012b) was designed. In this scale, respondents have to rate how they perceive access to work-family policies in their organization and they are asked whether they can use them without reprisals (e.g. "If I used work-family support resources, it would have negative consequences for my career", with an inverse scoring system). *EWB* was measured using an adapted version of the reflective 4-item scale ($\alpha_1=0.962$, $\alpha_2=0.962$, $\alpha_3=0.974$, $\alpha_4=0.977$ for flexi-time, long paid and unpaid leaves, flexi-place and family support services, respectively) designed by Boshoff & Mels (2000) and Warr (1990). This scale was based on the respondents' perception of work stress, job satisfaction, motivation to perform his/her tasks and organizational commitment (e.g., "I often feel anxious and stressed inside and outside my workplace", with inverse scoring, or "My main satisfactions in life come primarily from my job"). Finally, respondents' perception of their own *job performance* was measured by asking them to rate their job performance. In line with the recommendations of Boshoff & Mels (2000), the item "My organization gets the better of me in terms of job performance" was included in the questionnaire (α was not applicable, since it was a 1-item construct).

*3.3. Methods*

Following the recommendations of Hair et al (2014), hypothesis testing was performed using a structural equation model based on the PLS-SEM approach. The methodology selected –more specifically, the data collection methods employed–, as well as the constructs tested and the indicators used were appropriate for empirically examining the correlations among theoretical variables related to organizational work-family support (Casper et al, 2007; Chang et al, 2010). Data analysis was performed using SmartPLS 3.0 software (Ringle et al, 2014) and mean values were attributed to missing data using the criterion of replacement with average value. Although the parameters of the measurement model and the structural model were measured in a single step, the recommendations of Chin (2010) and Hair et al (2014) for the presentation of results were adopted. Accordingly, measurement model testing was performed first, followed by the evaluation of significance among parameters. As such, the validity and reliability of measurements was guaranteed before any conclusions were drawn on the relationships among constructs.

*3.4. Measurement model testing*

In this section, we examine whether the study variables (or indicators) measured the theoretical concepts correctly. Given that all constructs were reflective, reliability was analysed first; i.e., whether the indicators actually measured what they were intended to measure. Next, validity was evaluated; i.e., whether measuring was consistently performed.

In the reliability analysis, the reliability of each item was examined separately by assessing factor loadings (lambda). For an indicator to be definitely included in the measurement model of a construct, it must have a factor loading >=0.707. This involves the shared variance between the construct and its indicators being greater than the error variance. Some authors consider that this empirical rule ($\lambda >= 0.707$) should not be so rigid in early stages of scale design (Hair et al, 2014) and conclude that an indicator with a factor loading ranging from 0.4 to 0.7 can be deleted from a scale if deletion results in the average variance extracted (AVE) or Composite Reliability (CR) exceeding the minimum threshold value established (AVE = 0.5; CR = 0.7). Consequently, weak indicators can be occasionally maintained for their contribution to the validity of the content analysis measure. In any case, clearly weak indicators (≤0.4) must always be discarded.

In this study, all indicators of the measurement model were maintained, although two indicators did not reach the minimum threshold value established ($\lambda >= 0.707$). This decision was based on the fact that the AVE for all latent variables exceeded 0.5, which means that it was not necessary to delete these variables to reach the minimum threshold value for AVE. Therefore, as these two indicators contribute to the validity of the content, they were maintained.

The reliability of the scale was assessed to verify the internal consistency of all indicators when measuring the concept. Scale reliability was evaluated using Cronbach's Alpha coefficient and Composite Reliability (Table 2). Nunnally (1978) considered 0.7 adequate for indicating modest reliability and a stricter 0.8 for basic research. As shown in Table 2, all constructs comfortably exceeded the threshold established for *Cronbach's* Alpha and *Composite Reliability,* all except *Cronbach's* Alpha for the variable ACCESS 2, which was 0.792 and would be considered acceptable.

Construct validity was assessed by examining convergent validity and discriminant validity. *Convergent validity* signifies that a set of indicators represents one and the same underlying construct, which can be demonstrated through their unidimensionality (Henseler et al, 2009). Convergent validity is assessed through the average variance extracted (AVE), which provides a measure of the proportion of variance that can be explained by its indicators with respect to variance accounted for by measurement errors. Fornell & Lacker (1981) recommend a threshold value for AVE > 0.5, which means that 50% of the construct variance can be explained by its indicators but not by the indicators of the other constructs. As shown in Table 2, an AVE >0.5 was obtained for all constructs. This means that more than 50% of the variance in the construct can be accounted for by its indicators.

*Discriminant validity* examines to what extent a given construct differs from other constructs. Discriminant validity was assessed using the Fornell-Larcker criterion (1981), which is based on the idea that, in a given model, a construct should share more variance with its indicators than with other constructs. Consequently, an effective method for assessing discriminant validity is demonstrating that the AVE for a construct is greater than the variance that the construct shares with other constructs of the same model; in other words, the correlations among constructs are lower than the square of the AVE.

Table 2 shows in bold that the square of the AVE for all latent variables was greater than the correlation among variables. This means that all constructs were more strongly correlated to their indicators than to those of the other constructs.

Table 2. Evaluation of the measurement model

| GROUP 1 | Cronbach´s Alpha | Composite Reliability | AVE | ACCESSIBILITY1 | EWB1 | PERFORMANCE1 | EXISTENCE1 |
|---|---|---|---|---|---|---|---|
| ACCESSIBILITY1 | 0.819 | 0.917 | 0.846 | **0.920** | | | |
| EWB1 | 0.962 | 0.972 | 0.898 | 0.306 | **0.947** | | |
| PERFORMANCE1 | 1.000 | 1.000 | 1.000 | 0.251 | 0.892 | **1.000** | |
| EXISTENCE1 | 0.836 | 0.885 | 0.608 | 0.701 | 0.317 | 0.282 | **0.780** |
| GROUP 2 | Cronbach´s Alpha | Composite Reliability | AVE | ACCESSIBILITY2 | EWB2 | PERFORMANCE2 | EXISTENCE2 |
| ACCESSIBILITY2 | 0.792 | 0.906 | 0.828 | **0.910** | | | |
| EWB2 | 0.962 | 0.973 | 0.900 | 0.330 | **0.948** | | |
| PERFORMANCE2 | 1.000 | 1.000 | 1.000 | 0.283 | 0.899 | **1.000** | |
| EXISTENCE2 | 0.800 | 0.864 | 0.564 | 0.674 | 0.324 | 0.286 | **0.751** |
| GROUP 3 | Cronbach´s Alpha | Composite Reliability | AVE | ACCESSIBILITY3 | EWB3 | PERFORMANCE3 | EXISTENCE3 |
| ACCESSIBILITY3 | 0.851 | 0.930 | 0.870 | **0.933** | | | |
| EWB3 | 0.974 | 0.981 | 0.929 | 0.240 | **0.964** | | |
| PERFORMANCE3 | 1.000 | 1.000 | 1.000 | 0.228 | 0.922 | **1.000** | |
| EXISTENCE3 | 0.884 | 0.915 | 0.683 | 0.731 | 0.219 | 0.210 | **0.826** |
| GROUP 4 | Cronbach´s Alpha | Composite Reliability | AVE | ACCESSIBILITY4 | EWB4 | PERFORMANCE4 | EXISTENCE4 |
| ACCESSIBILITY4 | 0.885 | 0.945 | 0.897 | **0.947** | | | |
| EWB4 | 0.977 | 0.983 | 0.936 | 0.435 | **0.967** | | |
| PERFORMANCE4 | 1.000 | 1.000 | 1.000 | 0.380 | 0.922 | **1.000** | |
| EXISTENCE4 | 0.924 | 0.943 | 0.767 | 0.785 | 0.312 | 0.282 | **0.876** |

## 4. Results

Once the goodness of fit of the measurement model was verified, the structural model was analysed by assessing significance of correlation among variables. More specifically, the variance explained by endogenous variables was measured according to their $R^2$, path coefficients or standardized regression weights (Beta) and respective levels of significance. The $R^2$ of dependent latent variables provides a measure of the predictive power of a model and indicates the amount of variance of the construct that is accounted for by the model. Chin (1998) established the following threshold values for $R^2$: > 0.67 indicates "substantial"; > 0.33 indicates "moderate" and > 0.19 means a "weak" predictive value. The $R^2$ values obtained for this model show that the variable well-being has a weak predictive power ($R^2$: 0.114, 0.128, 0.062 and 0.1912 for each group of work-family policies, respectively), whereas the predictive power of the variable "Job performance" was substantial ($R^2$: 0.797, 0.809, 0.851 and 0.851 for flexi-time, long paid and unpaid leaves, flexi-place, and family support services, respectively. respectively).

The weak predictive power of the model for the variable well-being is due to the fact that apart from the existence of and access to work-family policies, this variable is affected by many other factors. These independent factors, however, were not the object of this study, and well-being, as it was measured in this study, accounts for an important proportion of variance in job performance.

In relation to path coefficients, they were calculated by bootstrapping. In accordance with standard guidelines (Hair et al, 2014), subsamples were created using 5,000 observations. Table 3 displays the *p*-value and confidence interval obtained for each path coefficient.

Table 3. Results of the structural model

| Hypotheses | GROUP 1 | Path coefficients | t-value (bootstrap) | P Values | Confidence Interval Low | Confidence Interval Up | Support | Observations |
|---|---|---|---|---|---|---|---|---|
| H1.1 | EXISTENCE1 -> PERFORMANCE1 | 0.028 | 1.426 | 0.154 | -0.010 | 0.068 | NO | |
| H2.1 | ACCESSIBILITY1 -> PERFORMANCE1 | -0.042 | 2.434 | 0.015 | -0.077 | -0.009 | NO | A positive effect was expected |
| H3.1 | EXISTENCE1 -> EWB1 | 0.201 | 5.383 | 0.000 | 0.128 | 0.277 | YES | |
| H4.1 | ACCESSIBILITY1 -> EWB1 | 0.165 | 4.532 | 0.000 | 0.092 | 0.232 | YES | |
| H5.1 | EWB1 -> PERFORMANCE1 | 0.896 | 87.737 | 0.000 | 0.876 | 0.915 | YES | |
| **Hypotheses** | **GROUP 2** | | | | | | | |
| H1.2 | EXISTENCE2 -> PERFORMANCE2 | 0.006 | 0.363 | 0.717 | -0.027 | 0.039 | NO | |
| H2.2 | ACCESSIBILITY2 -> PERFORMANCE2 | -0.020 | 1.207 | 0.227 | -0.052 | 0.013 | NO | |
| H3.2 | EXISTENCE2 -> EWB2 | 0.187 | 4.693 | 0.000 | 0.112 | 0.269 | YES | |
| H4.2 | ACCESSIBILITY2 -> EWB2 | 0.205 | 5.305 | 0.000 | 0.126 | 0.277 | YES | |
| H5.2 | EWB2 -> PERFORMANCE2 | 0.904 | 82.833 | 0.000 | 0.882 | 0.924 | YES | |
| **Hypotheses** | **GROUP 3** | | | | | | | |
| H1.3 | EXISTENCE3 -> PERFORMANCE3 | 0.007 | 0.446 | 0.656 | -0.023 | 0.038 | NO | |
| H2.3 | ACCESSIBILITY3 -> PERFORMANCE3 | 0.002 | 0.137 | 0.891 | -0.027 | 0.032 | NO | |
| H3.3 | EXISTENCE3 -> EWB3 | 0.093 | 2.637 | 0.008 | 0.024 | 0.162 | YES | |
| H4.3 | ACCESSIBILITY3 -> EWB3 | 0.173 | 4.897 | 0.000 | 0.103 | 0.240 | YES | |
| H5.3 | EWB3 -> PERFORMANCE3 | 0.920 | 109.027 | 0.000 | 0.903 | 0.936 | YES | |

| Hypotheses | GROUP 4 | | | | | | |
|---|---|---|---|---|---|---|---|
| H1.4 | EXISTENCE4 -> PERFORMANCE4 | 0.030 | 1.649 | 0.099 | -0.004 | 0.067 | NO | |
| H2.4 | ACCESSIBILITY4 -> PERFORMANCE4 | -0.051 | 2.426 | 0.015 | -0.093 | -0.010 | NO | A positive effect was expected |
| H3.4 | EXISTENCE4 -> EWB4 | -0.076 | 1.993 | 0.046 | -0.148 | 0.001 | NO | |
| H4.4 | ACCESSIBILITY4 -> EWB4 | 0.495 | 11.970 | 0.000 | 0.411 | 0.572 | YES | |
| H5.4 | EWB4 -> PERFORMANCE4 | 0.935 | 86.355 | 0.000 | 0.912 | 0.954 | YES | |

For n = 5,000 samples: * p < .05; ** p < .01; ***p < .001 (based on one-tailed Student's *t* test (4999)). t(0.05; 4999) = 1,645 ; t(0.01; 4999) = 2,327 ; t(0.001; 4999) = 3.092

Additionally, the same procedure was performed to assess the potential total and indirect effects among latent variables (Table 4 and Table 5).

Table 4. Indirect effects

| GROUP 1 | Path coefficients | t-value (bootstrap) | P Values | Confidence Interval Low | Confidence Interval Up | Statistically significant values |
|---|---|---|---|---|---|---|
| EXISTENCE1 -> PERFORMANCE1 | 0.180 | 5.368 | 0.000 | 0.115 | 0.249 | YES |
| ACCESSIBILITY1 -> PERFORMANCE1 | 0.148 | 4.488 | 0.000 | 0.082 | 0.209 | YES |
| **GROUP 2** | | | | | | |
| EXISTENCE2 -> PERFORMANCE2 | 0.169 | 4.680 | 0.000 | 0.101 | 0.242 | YES |
| ACCESSIBILITY2 -> PERFORMANCE2 | 0.185 | 5.263 | 0.000 | 0.114 | 0.250 | YES |
| **GROUP 3** | | | | | | |
| EXISTENCE3 -> PERFORMANCE3 | 0.085 | 2.634 | 0.008 | 0.022 | 0.149 | YES |
| ACCESSIBILITY3 -> PERFORMANCE3 | 0.159 | 4.865 | 0.000 | 0.094 | 0.221 | YES |
| **GROUP 4** | | | | | | |
| EXISTENCE4 -> PERFORMANCE4 | -0.071 | 1.992 | 0.046 | -0.139 | 0.001 | NO |
| ACCESSIBILITY4 -> PERFORMANCE4 | 0.463 | 11.654 | 0.000 | 0.382 | 0.538 | YES |

For n = 5,000 samples: * p < .05; ** p < .01; ***p < .001 (based on one-tailed Student's *t* test (4999)). t(0.05; 4999) = 1,645 ; t(0.01; 4999) = 2,327 ; t(0.001; 4999) = 3.092

Table 5. Total effects

| GROUP 1 | Path coefficients | t-value (bootstrap) | P Values | Confidence Interval Low | Confidence Interval Up | Statistically significant values |
|---|---|---|---|---|---|---|
| EXISTENCE1 -> PERFORMANCE1 | 0.208 | 5.516 | 0.000 | 0.136 | 0.283 | YES |
| ACCESSIBILITY1 -> PERFORMANCE1 | 0.105 | 2.918 | 0.004 | 0.032 | 0.173 | YES |
| EXISTENCE1 -> EWB1 | 0.201 | 5.383 | 0.000 | 0.128 | 0.277 | YES |
| ACCESSIBILITY1 -> EWB1 | 0.165 | 4.532 | 0.000 | 0.092 | 0.232 | YES |
| EWB1 -> PERFORMANCE1 | 0.896 | 87.737 | 0.000 | 0.876 | 0.915 | YES |
| **GROUP 2** | | | | | | |
| EXISTENCE2 -> PERFORMANCE2 | 0.175 | 4.366 | 0.000 | 0.100 | 0.256 | YES |
| ACCESSIBILITY2 -> PERFORMANCE2 | 0.165 | 4.244 | 0.000 | 0.086 | 0.240 | YES |
| EXISTENCE2 -> EWB2 | 0.187 | 4.693 | 0.000 | 0.112 | 0.269 | YES |
| ACCESSIBILITY2 -> EWB2 | 0.205 | 5.305 | 0.000 | 0.126 | 0.277 | YES |
| EWB2 -> PERFORMANCE2 | 0.904 | 82.833 | 0.000 | 0.882 | 0.924 | YES |
| **GROUP 3** | | | | | | |
| EXISTENCE3 -> PERFORMANCE3 | 0.092 | 2.626 | 0.009 | 0.025 | 0.164 | YES |
| ACCESSIBILITY3 -> PERFORMANCE3 | 0.161 | 4.556 | 0.000 | 0.090 | 0.229 | YES |
| EXISTENCE3 -> EWB3 | 0.093 | 2.637 | 0.008 | 0.024 | 0.162 | YES |
| ACCESSIBILITY3 -> EWB3 | 0.173 | 4.897 | 0.000 | 0.103 | 0.240 | YES |
| EWB3 -> PERFORMANCE3 | 0.920 | 109.027 | 0.000 | 0.903 | 0.936 | YES |

| GROUP 4 | | | | | | |
|---|---|---|---|---|---|---|
| EXISTENCE4 -> PERFORMANCE4 | -0.041 | 1.034 | 0.301 | -0.115 | 0.040 | NO |
| ACCESSIBILITY4 -> PERFORMANCE4 | 0.412 | 9.536 | 0.000 | 0.325 | 0.495 | YES |
| EXISTENCE4 -> EWB4 | -0.076 | 1.993 | 0.046 | -0.148 | 0.001 | NO |
| ACCESSIBILITY4 -> EWB4 | 0.495 | 11.970 | 0.000 | 0.411 | 0.572 | YES |
| EWB4 -> PERFORMANCE4 | 0.935 | 86.355 | 0.000 | 0.912 | 0.954 | YES |

For n = 5,000 samples: * p < .05; ** p < .01; ***p < .001 (based on one-tailed Student's t test (4999)). t(0.05; 4999) = 1,645 ; t(0.01; 4999) = 2,327 ; t(0.001; 4999) = 3.092

Table 6 includes a summary of the results obtained.

Table 6. Hypothesis testing

| Hypotheses | Description / type of support | Group 1 WFP for working hours flexibility (i=1) | | Group 2 WFP for flexibility in long leaves (i=2) | | Group 3 WFP for flexibility in location of work (i=3) | | Group 4 Family support services (i=4) | |
|---|---|---|---|---|---|---|---|---|---|
| | | Direct | Indirect | Direct | Indirect | Direct | Indirect | Direct | Indirect |
| H1.i | Positive correlation between WFP existence and job performance | NO | YES | NO | YES | NO | YES | NO | NO |
| H2.i | Positive correlation between WFP accessibility and job performance | NO | YES | NO | YES | NO | YES | NO | YES |
| H3.i | Positive correlation between WFP existence and EWB | YES | - | YES | - | YES | - | NO | - |
| H4.i | Positive correlation between WFP accessibility and EWB | YES | - | YES | - | YES | - | YES | - |
| H5.i | Positive correlation between EWB and job performance | YES | - | YES | - | YES | - | YES | - |

The results obtained do not support Hypothesis 1, which postulates that there may be a positive relationship between the existence of work-family policies and job performance. Conversely, the results do demonstrate that the existence of work-family policies based on flexibility in working time, long paid and unpaid leaves, and the work location has a total positive indirect effect on performance mediated by EWB. In contrast, the same does not occur to employee and family support services.

Hypothesis 2, which states that access to work-family policies may be positively related to job performance, is also not confirmed. Nevertheless, true access to work-family policies was found to have a total positive indirect effect on job performance mediated by EWB.

Hypotheses 1 and 2 suggest that EWB has a complete (rather than partial) mediating effect. The reason is that the relation between the existence of and access to work-family policies and job performance as mediated by EWB is significant, but the direct relationship between work-family policies existence and accessibility and job performance disappears when EWB is included as a mediating variable.

Hypothesis 3 posited that the existence of work-family policies may be positively related to EWB. The results obtained confirm this hypothesis for all work-family policies groups –except for employee and family support services (Group 4)– were proven to have a slight, statistically significant effect on job performance.

Hypothesis 4 suggested that work-family policies accessibility may be positively related to EWB. This hypothesis is confirmed for all work-family policies groups, as they were found to have a slight but statistically significant effect on job performance.

Finally, hypothesis 5 is also confirmed by the results of the study. Thus, EWB has a significant effect on improvement in job performance. EWB was the only variable found to have a significant, direct effect on job performance.

## 5. Discussion and implications for human resource management

In accordance with the results obtained in this study, the existence of and access to work-family policies do not have a direct effect on job performance, but rather an indirect one when mediated by the EWB generated by work-family policies. These results extend the analysis of Steenbergen & Ellemers (2009) of the banking sector, who reported that work-family policies are positively related to physical EWB and job performance. The authors, however, did not consider the mediating role of EWB in the work-family policies-performance relationship. Furthermore, their study examined the impact of work-family policies on physical EWB based on health indicators. In contrast, our study focused on emotion-based EWB, which can be considered a precursor of physical EWB, work stress being the catalyst for emotional EWB degenerating in physical distress (van Steenbergen & Ellemers, 2009).

On the other hand, our results demonstrate that the existence of and access to some types of work-family policies (see Table 1) are indirectly related to job performance. This relationship has been proven to exist in all types of work-family policies except for employee and family support services (Group 4 in Table 1). In light of the above, it can be deduced that employees do not appreciate, for instance, the existence of employee and family support services for carers of children and elderly relatives as much as other work-family policies related to working time and workplace flexibility. This finding contrasts the high proportion of respondents who had child or elder care responsibilities (75.9%). Conversely, respondents did appreciate that the work-family policies available were actually accessible and did not involve any reprisals.

Another relevant aspect to take into account is the context where this study was carried out. It is widely known that the economic and financial crisis in Spain negatively affected labour rights –and especially work-family policies– in this sector (Alamá et al, 2015; Mihelič & Tekavčič, 2014; Maudas, 2011). This was added to work stress and the pressure traditionally exerted on employees for the improvement of results in this sector (Burke, 2009; Rosso, 2008). In this context, it seems that employees were more concerned about maintaining their job than on their own well-being. The results of this study, however, contradict this assumption and demonstrate that EWB is positively related to job performance.

The application of the findings of this study to the sphere of human resource management is threefold. First, HR managers can improve EWB by adopting work-family policies, informing on their availability, and ensuring they can be used without reprisals. Thus, the to-do list of any company should include a supportive work-family organizational culture that promotes organizational and supervisor support to employees with family responsibilities (Las Heras et al, 2015; Sivatte & Guadamillas, 2014).

Second, the EWB generated by the implementation of work-family policies may have a positive effect on job performance and improve the cost-benefit ratio. Analysing the cost-benefit ratio is helpful for managers, as the impact of work-family policies –and many other policies– on the organization is generally assessed in terms of cost-benefit (McDonald et al, 2005).

Third, as stated above, managers should be aware that job performance will improve more substantially if flexi-time, long leaves, and flexi-place arrangements –rather than employee and family support services– are offered to their employees.

Finally, it should be noted that HR managers must take into account that work-family policies are not only important for employees with family responsibilities, but for all employees. Although work-family policies are aimed at this type of employee, increasing attention is being paid to employees without family responsibilities who want to maintain a work-life balance (Felstead et al, 2002).

## 6. Conclusions

This study makes a theoretical and empirical contribution to better understand the impact that the existence of and true access to work-family policies have on job performance, mediated by EWB. More specifically, this study assesses the effects of different types of work-family policies on EWB and job performance. The first group of work-family policies encompasses working time flexibility policies (flexi-time). The second group consists of long paid and unpaid leave policies. The third group includes policies providing flexibility in the work location (flexi-place). Finally, the fourth group embraces employees and family support services.

Work-family policies were evaluated from two perspectives: (1) the existence and employees' awareness of work-family policies; and (2) true access to work-family policies. The results obtained indicate that the existence of and access to work-family policies do not have a direct effect on job performance but an indirect one mediated by EWB. This was found to be applicable to all groups except for employee and family support services. The existence of this type of work-family policies does not seem to have a direct or indirect effect on job performance or EWB. Yet, once this type of work-family policies has been adopted, it is important that employees perceive they can use them easily, as this has an indirect positive effect on job performance.

A limitation of this study is that fieldwork focused on a single sector –the banking sector– and on a single country –Spain–. Thus, legal and cultural differences may hinder the extrapolation of the results obtained to other sectors and countries (Allen et al, 2015).

In future research, work-family policies should be analysed in other sectors and countries, since context may cause significant differences in terms of WFC (Allen et al, 2015) and work-family policies.

Furthermore, other variables related to EWB should be included in future studies as model control variables. In this line, demographic, socioeconomic, sectorial, geographical, cultural variables should be also considered, among others.

Given the impact of work-family policies on EWB and job performance, further studies should be conducted on the academic and practical implications of these variables for human resource management.